\documentclass[a4paper,12pt]{article}
\usepackage{amsmath,amsthm,amssymb,bbm}

\usepackage[english]{babel}

\title{\bf Environment induced entanglement\\ in a refined weak-coupling limit}

\author{F. Benatti$^{a,b}$,
R. Floreanini$^{b}$ and U. Marzolino$^{a,b}$\\
\small ${}^a$Dipartimento di Fisica Teorica, Universit\`a di Trieste,
34014 Trieste, Italy\\
\small ${}^b$Istituto Nazionale di Fisica Nucleare, Sezione di Trieste,
34014 Trieste, Italy}

\date{\null}

\begin{document}

\maketitle

\begin{abstract}
\noindent
Two non-directly interacting qubits with equal frequencies can become entangled
via a Markovian, dissipative dynamics through the action of a weakly coupled Ohmic heat bath.
In the standard weak-coupling limit derivation, this purely dissipative effect disappears if the frequencies are different because of the ``ergodic average'' used by this approach. However,
there are physical situations where this technique is too rough to capture all the relevant aspects of the dissipative dynamics.
In these cases, in order to better describe the physical behavior of the open system, it is necessary to go beyond
the ``ergodic average''.
We show that, in this more refined framework, the entanglement capability of the
environment persists also in the case of different frequencies.
\end{abstract}

Quantum systems are usually treated as isolated: this is just an approximation,
justified for vanishingly small couplings with the external environment.
When the interaction with the environment
is weak but not negligible, a reduced dynamics can be obtained by eliminating the
environment degrees of freedom and by subsequently performing a so-called Markovian approximation~\cite{Spohn}-\cite{BenattiFloreanini3}.
These systems are known as open quantum systems and their reduced dynamics is irreversible and satisfies a forward-in-time composition law: it is described by a so-called quantum dynamical semigroup that incorporates the dissipative and noisy effects due to the environment.
The latter acts as a source of decoherence: in general, the corresponding reduced dynamics irreversibly transforms pure states (one-dimensional projections) into statistical mixtures (density matrices).

One of the most intriguing aspects of quantum coherence is entanglement~\cite{Horo}, that is the existence of purely quantum mechanical correlations, which has become a central topic in quantum information for its many applications as a physical resource enabling otherwise impossible information processing protocols.
With reference to the entanglement content of a state of two qubits
embedded in a same heat bath, it is generally expected that it would be depleted by decoherence effects.
However, this is not the only possibility: if suitably engineered, the environment can entangle an initial separable state of two dynamically independent systems; the reason is that, although not directly interacting between themselves, there can be an environment mediated generation of quantum correlations between two systems immersed in it.

This possibility has been demonstrated analytically for two qubits with a same oscillation frequency~\cite{Braun,BenattiFloreanini1,Bose} and two identical harmonic oscillators~\cite{BenattiFloreanini2}
evolving according to a reduced master equation of the typical Lindblad form~\cite{GoriniKossakowskiSudarshan,Lindblad}, obtained via the so-called weak-coupling limit.~\footnote{For the system composed by two harmonic oscillators immersed in a heat bath of other oscillators
similar results have been obtained by studying numerically their exact time evolution \cite{PazRoncaglia}.}
This technique is based on the fact that the dissipative effects are visible only over a coarse-grained time-scale $\Delta t$, so large that the free dynamics of the embedded systems can be averaged out over $\Delta t$~\cite{AlickiLendi}-\cite{BenattiFloreanini3}. The resulting elimination of too rapid oscillations
is mathematically implemented through a time ``ergodic average''.

Noticeably, such a prescription guarantees that, unlike for reduced dynamics of Redfield type (see~\cite{DumckeSpohn,BenattiFloreanini3}), the resulting quantum dynamical semigroups consist of
completely positive maps~\cite{Spohn,AlickiLendi,Davies,GoriniKossakowskiSudarshan}.
Complete positivity ensures that the open quantum evolution is consistent with entanglement in the sense that not only the positivity of any initial density matrix of the open system is preserved in time, but also that of any initial state of the open system coupled to any other possible ancillary system.
Indeed, only complete positivity can guarantee the full physical consistency of the Markovian
approximations that one uses to describe an open quantum dynamics; in other words, without
complete positivity, it always occurs that at least one initial state carrying entanglement between
the given system and an ancilla will assume negative eigenvalues in the course of time~\cite{BenattiFloreanini3}.

It turns out that, in the standard weak-coupling limit approximation, when two qubits  or two harmonic oscillators embedded in a same environment have different oscillation frequencies $\omega_1\neq\omega_2$, no matter how small the difference $\omega_2-\omega_1$ is, the elimination of too rapid oscillations destroys the generation capability that the environment possesses when $\omega_1=\omega_2$.

In this paper, we study this behavior in the case of two qubits weakly interacting, via a Ohmic coupling, with a heat bath made of free bosons at high temperature. We shall first relate the sharp dependence of the entanglement capability of the environment to the oscillation frequencies of the two qubits on the drastic elimination of too fast oscillations through the ``ergodic average''.
This procedure is only allowed when the coupling to the environment is such that the coarse-grained time-scale $\Delta t$ can effectively be considered infinitely large. However, there are situations where this approximation is not physically meaningful and $\Delta t$ should be kept finite. In these cases, to better capture the effects of the open system dynamics, a less rough time coarse-graining is needed; in the following, we use a master equation derived without recourse to the ergodic average that nevertheless generates a completely positive dynamical semigroup.
We show that, in this refined framework, the entanglement generation capability of the environment is preserved even when $\omega_1\neq \omega_2$.

\vskip 1cm

The problem we will address in the following regards whether two non-interacting qubits
with Hamiltonian
\begin{equation}
H_S=\frac{\omega_1}{2}\,\sigma^{(1)}_3\,+\,\frac{\omega_2}{2}\,\sigma^{(2)}_3\ ,
\end{equation}
can become entangled when weakly coupled to free spinless Bosons in thermal equilibrium
via a (finite volume) interaction of the form
\begin{equation}
H_I=\lambda\,\,X(f)\,\Bigl(\sigma^{(1)}_1\,+\,\sigma^{(2)}_1\Bigr)\ ,\qquad
X(f)=\sum_k\Bigl(f(k)\,a^\dag_k+f^*(k)\,a_k\Bigr)\ ,
\end{equation}
where $a^\dag_k$ and $a_k$ denote the creation
and annihilation operators of Bose modes with momentum $k$ and energy $\omega(k)$, $f(k)$ is a one-particle
Bose state in momentum representation, $\lambda$ is a small, dimensionless coupling constant, while
$\sigma^{(1)}_{1,3}=\sigma_{1,3}\otimes 1$ and $\sigma^{(2)}_{1,3}=1\otimes\sigma_{1,3}$
represent the first and third Pauli matrices for the two qubits.~\footnote{
For sake of simplicity, we consider the two qubits located at a same point in space.
For the effects on entanglement creation that result when the two qubits are spatially separated we refer
to~\cite{BenattiFloreanini1,Bose} for equal frequencies and to~\cite{BFMprep} in the case of $\omega_1\neq\omega_2$.}
The total system  comprising the two qubits and the thermal bosons will thus be described
by the Hamiltonian
\begin{equation}
H=H_S\,+\,H_B\,+\,H_I\ ,\qquad  H_B=\sum_k\,\omega(k)\,a^\dag_k\,a_k.
\end{equation}

In general, the state of the compound system $S+B$ at time $t>0$ is a correlated state $\rho_{SB}(t)$
from which the state of the two qubits can be extracted as $\rho(t)={\rm Tr}_B(\rho_{SB}(t))$ by tracing out the environment degrees of freedom; however, the time-evolution equation for $\rho(t)$ is quite complicated.
In order to arrive at a memoryless master equation, one starts with the physically
acceptable hypothesis that the initial state of open system plus environment be of the uncorrelated form $\rho\otimes\rho_\beta$, where $\rho_\beta\propto\exp{(-\beta H_B)}$
is the Boson thermal equilibrium state at inverse temperature $\beta$.
Then, one goes to the interaction representation by replacing $\rho(t)$ with
$\widetilde{\rho}_{SB}(t)={\rm e}^{it(H_S+H_B)}\,\rho_{SB}(t)\,{\rm e}^{-it(H_S+H_B)}$
and looks at the time-evolution over time-intervals $\Delta t$ by stopping at the first significative order in
$\lambda$ in the time-ordered expansion of $\widetilde{\rho}(t+\Delta t)$. Then,
\begin{equation}
\frac{\widetilde{\rho}(t+\Delta t)-\widetilde{\rho}(t)}{\Delta t}\simeq
-\frac{\lambda^2}{\Delta t}\int_t^{t+\Delta t}{\rm d}t_1\int_t^{t_1}{\rm d}t_2
{\rm Tr}\Biggl(
\Bigl[H_I(t_1),\Bigl[H_I(t_2),\widetilde{\rho}_{SB}(t)\Bigr]\Bigr]\Biggr)\ ,
\label{APP1}
\end{equation}
where
\begin{equation}
H_I(t)=X_t(f)\, \Bigl(\sigma^{(1)}_1(t)+\sigma^{(2)}_1(t)\Bigr)\ ,\qquad
X_t(f)=\sum_k\Bigl(f(k){\rm e}^{it\omega(k)}a^\dag_k+f^*(k){\rm e}^{-it\omega(k)}a_k\Bigr)\ ,
\end{equation}
with
$\sigma_1^{(a)}(t)=\cos(\omega_at)\,\sigma^{(a)}_1-\sin(\omega_at)\,\sigma^{(a)}_2$, $a=1,2$.

Notice that the variation of $\widetilde{\rho}(t)$ is of order $\lambda^2$, namely it is non-negligible
only over times $t=\tau/\lambda^2$;
if $\lambda\ll 1$ and $\Delta t$ is such that $\lambda^2\Delta t$ is small on the scale of $\tau$,
one may reasonably substitute in~(\ref{APP1}) the finite ratio with a time-derivative:
\begin{equation}
\label{APP4}
\frac{\widetilde{\rho}\big((\tau+\lambda^2\Delta t)/\lambda^2\big)
-\widetilde{\rho}\big(\tau/\lambda^2\big)}{\Delta t}\simeq\partial_t\widetilde{\rho}(t)\ .
\end{equation}
Indeed, the error is bounded by $\lambda^2\Delta t$.
Further, if $\Delta t$ is chosen much larger than the decay time $\tau_B$
of the environment two-point time-correlation functions, one may approximate $\widetilde{\rho}_{SB}(t)$
with the uncorrelated state
$\rho(t)\otimes\rho_\beta$ as it was at $t=0$~\cite{AlickiLendi,BreuerPetruccione}.
Therefore, if $\tau_B\ll\Delta t$, one gets the following approximated
master equation (in interaction representation):
\begin{equation}
\partial_t\widetilde{\rho}(t)=
-\frac{\lambda^2}{\Delta t}\int_t^{t+\Delta t}{\rm d}t_1\int_t^{t_1}{\rm d}t_2{\rm Tr}\Biggl(
\Bigl[H_I(t_1)\,,\,\Bigl[H_I(t_2),\,\,\widetilde{\rho}(t)\otimes\rho_\beta\Bigr]\Bigr]\Biggr)\ .
\label{APP5}
\end{equation}

The meaning of $\Delta t$ is that of a time-coarse graining parameter naturally associated
to the slow dissipative time-scale $\tau=t\lambda^2$. More precisely,
significant variations of the system density matrix due to the presence of the environment can only be seen after
a time $\Delta\tau=\lambda^2\Delta t$ has elapsed.
Given the coupling constant $\lambda\ll1$, the dissipative time-scale is set and
actual experiments cannot access faster time-scales; furthermore,
if $\Delta\tau\ll1$ on the scale of $\tau$, then the experimental evidences are consistently described by
a master equation as in~(\ref{APP5}).
The weak-coupling limit consists in letting $\lambda\to0$~\cite{AlickiLendi,BreuerPetruccione};
then, variations of the system density matrix
as in~(\ref{APP5}) are actually visible only if $\Delta t\to+\infty$.
This allows one to eliminate all fast oscillations through a time ``ergodic average'' as in the standard
weak-coupling limit approach~\cite{Davies,AlickiLendi}.
Instead, if the system-environment coupling $\lambda$ is small, but not vanishingly small, then
$\Delta t$ cannot be taken infinitely large. In such cases, the usual weak-coupling limit techniques provide an approximation which is too rough to properly describe the dissipative time-evolution and one needs a more refined approach in order to be able to keep contributions that would otherwise be washed out.
\footnote{A different approach is developed in \cite{Alicki}: there, a non-Markovian weak coupling approximation
of the reduced dynamics is introduced, leading to a two-parameter family
of dynamical maps, with a time-dependent generator \cite{AlickiLendi}.
We stress that this treatment is completely different from the one discussed below,
which instead describes the reduced two-atom dynamics in terms of a Markovian,
one parameter semigroup.}

Returning to the Schr\"odinger representation,~(\ref{APP5}) yields the
following memoryless master equation (see~\cite{Alicki,Lidar,Schaller}) of Kossakowski-Lindblad type~\cite{AlickiLendi,BreuerPetruccione}:
\begin{equation}
\label{ME0}
\partial_t\rho(t)=-i\Bigl[H_S+\lambda^2\,H_{\Delta t}\,,\,\rho(t)\Bigr]\,+\,D_{\Delta t}[\rho(t)]\ .
\end{equation}
The environment contributes to the generator of the reduced dynamics with an Hamiltonian $H_{\Delta t}$ and a
purely dissipative term $D_{\Delta t}[\rho(t)]$; both of them depend on the environment through the
two-point time-correlation functions
\begin{eqnarray}
 \label{corr-funct}
G(t)={\rm Tr}\Bigl(\rho_\beta X(f)\,X_t(f)\Bigr)=G(-t)^*\ .
\end{eqnarray}

The bath-induced Hamiltonian $H_{\Delta t}$ contains a bath-mediated qubit-qubit interaction
\begin{eqnarray}
\label{Hamcorr}
H^{int}_{\Delta t}&=&\sum_{i,j=1,2}h_{ij}(\Delta t)\,\sigma^{(1)}_i\sigma^{(2)}_j\ ,
\end{eqnarray}
where the $2\times 2$ matrix $h_{\Delta t}=[h_{ij}(\Delta t)]$ is real, but not necessarily Hermitian.
It is convenient to introduce the following matrices $\displaystyle
[\Psi_{\varepsilon j}]=\frac{1}{2}\begin{pmatrix}
1& i\cr 1&-i
\end{pmatrix}$, $\varepsilon=\pm 1$; then
\begin{equation}
h_{\Delta t}=\Psi^\dagger\,H^{(12)}_{\Delta t}\,\Psi\,+
\,\Bigl(\Psi^\dagger\,H_{\Delta t}^{(21)}\,\Psi\Bigr)^T\ ,
\end{equation}
where $H^{(ab)}_{\Delta t}=[H^{(ab)}_{\varepsilon\varepsilon'}(\Delta t)]$, $a,b=1,2$,
is explicitly given by
\begin{equation}
H^{(ab)}_{\varepsilon\varepsilon'}(\Delta t)=
-\frac{i}{2\Delta t}\int_0^{\Delta t}{\rm d}t_1\int_0^{\Delta t}{\rm d}t_2\
{\rm e}^{i(\varepsilon\omega_at_1-\varepsilon'\omega_bt_2)}\hbox{sign}(t_2-t_1)G(t_2-t_1)\ ,
\label{Ham12}
\end{equation}
where $X^T$ denotes matrix transposition. Instead, the purely dissipative part can be written as
\begin{equation}
D_{\Delta t}[\rho(t)]=
\sum_{a,b=1,2;\ i,j=1,2} C^{(ab)}_{ij}(\Delta t)\
\Biggl(
\sigma^{(a)}_i\,\rho(t)\,\sigma^{(b)}_j-\,\frac{1}{2}\Bigl\{\sigma^{(b)}_j
\sigma^{(a)}_i\,,\,\rho(t)\Bigr\}\Biggr)\ ,
\label{Disscorr}
\end{equation}
where the $2\times2$ matrices $C^{(ab)}_{\Delta t}=[C^{(ab)}_{ij}(\Delta t)]$ read
\begin{equation}
C^{(ab)}_{\Delta t}=\Psi^\dag D^{(ab)}_{\Delta t}\Psi\ ,
\end{equation}
in terms of the matrices
$D^{(ab)}_{\Delta t}=[D^{(ab)}_{\varepsilon \varepsilon'}(\Delta t)]$
with entries
\begin{equation}
D^{(ab)}_{\varepsilon\varepsilon'}(\Delta t)=
\frac{1}{\Delta t}\int_0^{\Delta t}{\rm d}t_1\int_0^{\Delta t}{\rm d}t_2\,
{\rm e}^{i(\varepsilon\omega_at_1-\varepsilon'\omega_bt_2)}\,G(t_2-t_1)\ .
\label{Diss}
\end{equation}
The $4\times 4$ Kossakowski matrix
$C_{\Delta t}$ formed by the $2\times 2$ blocks $C^{(ab)}_{\Delta t}$,
$a,b=1,2$,
involves the Fourier transforms of the
correlation functions~(\ref{corr-funct}) and is automatically positive definite; this fact guarantees that
the master equation~(\ref{ME0}) generates a semigroup of dynamical maps $\gamma_t^{\Delta t}$
on the two-quibit density matrices
which are completely positive~\cite{Spohn}-\cite{BenattiFloreanini3}, whence,
as discussed in the introduction, fully physically consistent.

Given two qubits weakly interacting with their environment, a sufficient condition for them
to get entangled at small times by the completely positive reduced dynamics has been derived in~\cite{BenattiFloreaniniPiani,BenattiLiguoriNagy} and is based on the properties of the generator
in~(\ref{ME0}), that is on the interaction Hamiltonian~(\ref{Hamcorr}) and the dissipative contribution~(\ref{Disscorr}).
We shall focus on an initial two qubit state of the form
$|\downarrow\rangle\otimes|\uparrow\rangle$, where
$|\downarrow\rangle$, $|\uparrow\rangle$ are the eigenstates of $\sigma_3$; then,
the condition is as follows:
\begin{equation}
\delta=D^{(11)}_{--}(\Delta t)\, D^{(22)}_{++}(\Delta t)-\left|{\cal D}^{(12)}_{\Delta t}
+i{\cal H}^{(12)}_{\Delta t}\right|^2<0\ ,
\label{suffcond2}
\end{equation}
where, from~(\ref{Ham12}) and~(\ref{Diss}),
\begin{eqnarray}
\label{suffcond3c}
{\cal D}_{\Delta t}^{(12)}&=&
\frac{D^{(12)}_{--}(\Delta t)+(D^{(12)}_{--}(\Delta t))^*}{2}\\
\label{suffcond3d}
{\cal H}^{(12)}_{\Delta t}&=&H^{(12)}_{--}(\Delta t)+H^{(21)}_{++}(\Delta t)\ .
\end{eqnarray}
If $\delta>0$, such a separable state
cannot get entangled by the environment at small times~\cite{BenattiLiguoriNagy}.

We are interested in the capacity of the environment to generate entanglement, at least at small times with respect to the dissipative time-scale. Because of the positivity of the Kossakowski matrix $C_{\Delta t}$ the diagonal block matrices $C^{(aa)}_{\Delta t}$ and $D_{\Delta t}^{(aa)}$, $a=1,2$, are positive definite; therefore, the dissipative entanglement generation depends on the quantity ${\cal D}_{\Delta t}^{(12)}$,
but also on how it interferes with ${\cal H}_{\Delta t}^{(12)}$ in~(\ref{suffcond2}).
In the case at hand, it turns out that
\begin{equation}
\left|{\cal D}_{\Delta t}^{(12)}\,+\,i\,{\cal H}_{\Delta t}^{(12)}\right|^2\,=\,
\left|{\cal D}_{\Delta t}^{(12)}\right|^2\,+\,\left|{\cal H}_{\Delta t}^{(12)}\right|^2\ .
\end{equation}
Indeed, using~(\ref{corr-funct}), one can check by explicit computation that the quantities
${\rm e}^{-i(\omega_2-\omega_1)\Delta t/2}\,{\cal D}_{\Delta t}^{(12)}$ and
${\rm e}^{-i(\omega_2-\omega_1)\Delta t/2}\,{\cal H}_{\Delta t}^{(12)}$
are both real.
We shall thus concentrate on the purely dissipative entanglement generation, that is on the difference
\begin{equation}
\label{diss-diff}
\tilde{\delta}=D^{(11)}_{--}(\Delta t)\,D_{++}^{(22)}(\Delta t)\,-\,\left|{\cal D}_{\Delta t}^{(12)}\right|^2\ .
\end{equation}
If $\tilde{\delta}$ is negative, {\it a fortiori} also $\delta$ in~(\ref{suffcond2}) is negative.

We shall consider (infinite volume) Ohmic correlation functions~\cite{Palma}
\begin{equation}
G(t)=\int_0^{+\infty}\hskip-.5cm
{\rm d}\omega\,{\rm e}^{-\omega/\omega_c}\,\omega\,
\Bigl(\coth\frac{\beta\omega}{2}\,\cos\omega t\,-\,i\,\sin\omega t\Bigr)\ ,
\end{equation}
where $\omega_c$ is a Debye cut-off;
then, setting $\displaystyle{\rm sinc}(x)=\frac{\sin x}{x}$ and $(a,\varepsilon)=(1,-),(2,+)$, one
explicitly gets

\begin{eqnarray}
\label{suffcond4a}
&&\hskip-1cm
D^{(aa)}_{\varepsilon\varepsilon}(\Delta t)=\frac{\varepsilon\Delta t}{2}
\int_{-\infty}^{+\infty}{\rm d}\omega\,\omega\,\frac{{\rm e}^{-|\omega|/\omega_c}}{{\rm e}^{\varepsilon\beta\omega}-1}\,
{\rm sinc}^2\bigg[\frac{(\omega-\omega_a)\Delta t}{2}\bigg]\ ,\\
&&\hskip-1cm
\Bigl|{\cal D}^{(12)}_{\Delta t}\Bigr|=\frac{\Delta t}{2}
\int_{-\infty}^{+\infty}{\rm d}\omega\,\omega\,{\rm e}^{-|\omega|/\omega_c}\,
\coth{(\beta\omega/2)}\,{\rm sinc}\bigg[\frac{(\omega-\omega_1)\Delta t}{2}\bigg]\,
{\rm sinc}\bigg[\frac{(\omega-\omega_2)\Delta t}{2}\bigg]\ .
\label{suffcond4c}
\end{eqnarray}

The weak-coupling limit amounts to $\Delta t\to+\infty$;
since $\displaystyle\lim_{\alpha\to+\infty}\frac{\alpha}{2\pi}{\rm sinc}(\alpha(x-x_0))=\delta(x-x_0)$, one finds
\begin{eqnarray}
\label{suffcond5a}
&&
\hskip-1cm
\lim_{\Delta t\to+\infty}D^{(aa)}_{\varepsilon\varepsilon}(\Delta t)=
4\pi\varepsilon\,\frac{\omega_a\,{\rm e}^{-\omega_a/\omega_c}}{{\rm e}^{\varepsilon\beta\omega_a}-1}\\
&&
\label{suffcond5c}
\hskip-1cm
\lim_{\Delta t\to+\infty}\Bigl|{\cal D}^{(12)}_{\Delta t}\Bigr|=2\pi\delta_{\omega_1\omega_2}\,
\omega_2\,{\rm e}^{-\omega_2/\omega_c}\coth\frac{\beta\omega_2}{2}\ .
\end{eqnarray}
If $\omega_1=\omega_2$,
the difference~(\ref{diss-diff}) is always negative:
$\displaystyle
\tilde{\delta}=-4\pi^2\,\omega_2^2\,{\rm e}^{-2\omega_2/\omega_c}$;
instead, if $\omega_1\neq\omega_2$,
$\displaystyle\tilde{\delta}=D^{(11)}_{--}(\Delta t)\,D^{(22)}_{++}(\Delta t)\geq 0$.
Thus, in the latter case, the initial separable state
$\vert \downarrow\rangle\otimes\vert\uparrow\rangle$
cannot get dissipatively entangled by the environment at small times, unless $\omega_1=\omega_2$.

This behavior is characteristic of physical situations where an infinitely large coarse-grained
time-scale $\Delta t$ is justified by a vanishingly small coupling $\lambda$ to the environment.
However, if $\lambda$ is small, but not negligibly small, the
qubit density matrix effectively varies over large but finite $\Delta t$. Then,
terms of order $1/\Delta t$ like
\begin{equation}
\label{formula}
\Bigl|{\cal D}_{\Delta t}^{(12)}\Bigr|\simeq \pi\,{\rm sinc}(\delta\omega\Delta t)\sum_{a=1}^2
\omega_a\,{\rm e}^{-\omega_a/\omega_c}\,\coth(\beta\omega_a/2)
\end{equation}
can make $\tilde{\delta}$ negative even when $\delta\omega=(\omega_2-\omega_1)/2>0$.
Indeed, for high temperatures $\beta\omega_{1,2}\ll 1$ and large cut-offs
$\omega_{1,2}/\omega_c\ll1$, expanding~(\ref{suffcond5a}) and~(\ref{formula})
yields~\footnote{The high temperature hypothesis has been made in order to permit an analytical study of the behavior of $\tilde{\delta}$. This is the worst possible scenario; indeed, in general, lower temperatures favor the entanglement creation capability of the environment~\cite{BFMprep}.}
\begin{equation}
\tilde{\delta}\simeq
\frac{16\pi^2}{\beta^2}\Bigl(1-\beta\delta\omega-{\rm sinc}^2(\delta\omega\Delta t)\Bigr)\ .
\end{equation}
If the qubit frequency difference $\delta\omega$ and the coupling strength $\lambda$ are
such that $\delta\omega\Delta t\ll1$ then a further expansion of the $\rm sinc$ function yields
\begin{equation}
\tilde{\delta}\simeq-\frac{16\pi^2}{\beta^2}\Bigl(\beta\delta\omega-\frac{(\delta\omega)^2(\Delta t)^2}{3}\Bigr)\ ,
\end{equation}
which is negative when
$\displaystyle \frac{1}{\omega_{1,2}}\gg\beta\,>\,\frac{\delta\omega(\Delta t)^2}{3}$.

In conclusion, the fact that the environment may generate two-qubit entanglement only if the qubit
frequencies are equal, is a consequence of the weak-coupling limit and the associated time ``ergodic average'';
in such a case, the coupling constant $\lambda\to 0$ so that the coarse-grained time-scale $\Delta t$ over which the qubit density matrix effectively changes due to the presence of the environment become so large that all off-resonant
phenomena are averaged out.
Instead, if the physical conditions ask for a coupling constant $\lambda$ that is small but not vanishingly so,
a consistent description of the open dynamics requires $\Delta t$ finite; in this way, one may keep track of finer effects and save the possibility of dissipative entanglement generation even if $\omega_1\neq \omega_2$.
\vfill
\eject

\end{document}